\documentclass[onecolumn,twoside]{ieeetran}
\usepackage{graphicx,subfigure}
\usepackage{amssymb}
\begin{document}

\title{DNN Filter Bank Cepstral Coefficients for Spoofing Detection}
\author{Hong~Yu,
        ~Zheng-Hua~Tan,~\IEEEmembership{Senior Member,~IEEE,}
        ~Zhanyu~Ma,~\IEEEmembership{Member,~IEEE,}
        and~Jun~Guo
\thanks{H. Yu, Z. Ma, and J. Guo are with Pattern Recognition and Intelligent
System Lab., Beijing University of Posts and Telecommunications, Beijing,
China.}
\thanks{Z.-H. Tan is with the Department of Electronic Systems, Aalborg University,
Aalborg, Denmark}
\thanks{This work was conducted during H. Yu's visit to Z.-H. Tan at the Aalborg University.}
\thanks{The corresponding author is Z. Ma. Email mazhanyu@bupt.edu.cn}

}

\maketitle

\begin{abstract}
With the development of speech synthesis techniques, automatic speaker verification systems face the serious challenge of spoofing attack. In order to improve the reliability of speaker verification systems, we develop a new filter bank based cepstral feature, deep neural network filter bank cepstral coefficients (DNN-FBCC), to distinguish between natural and spoofed speech. The deep neural network filter bank is automatically generated by training a filter bank neural network (FBNN) using natural and synthetic speech. By adding restrictions on the training rules, the learned weight matrix of FBNN is band-limited and sorted by frequency, similar to the normal filter bank. Unlike the manually designed filter bank, the learned filter bank has different filter shapes in different channels, which can capture the differences between natural and synthetic speech more effectively. The experimental results on the ASVspoof {2015} database show that the Gaussian mixture model maximum-likelihood (GMM-ML) classifier trained by the new feature performs better than the state-of-the-art linear frequency cepstral coefficients (LFCC) based classifier, especially on detecting unknown attacks.
\end{abstract}



\section{Introduction}
\IEEEPARstart{A}{s} a low-cost and flexible biometric solution to person authentication, automatic speaker verification (ASV) has been used in many telephone or network access control systems, such as telephone banking~\cite{wu2013synthetic}. Recently, with the improvement of automatic speech generation methods, speech produced by voice conversion (VC)~\cite{wu2013conditional}\cite{helander2012voice} and speech synthesis (SS)~\cite{hunt1996unit}\cite{ze2013statistical} techniques has been used to attack  ASV systems. Over the past few years, much research has been devoted to protect ASV systems against spoofing attack~\cite{sizov2015joint}\cite{tian2016spoofing}\cite{sanchez2015toward}.

There are two general strategies to protect ASV systems.
One is to develop a more robust ASV system which can resist the spoofing attack. Unfortunately, research has shown that all the existing ASV systems are vulnerable to spoofing attacks~\cite{kinnunen2012vulnerability}\cite{de2012evaluation}\cite{lindberg1999vulnerability}. Verification and anti-spoofing tasks can not be done well in only one system at the same time.

The other more popular strategy is to build a separated spoofing detection system which only focuses on distinguishing between natural and synthetic speech~\cite{sahidullahintegrated}. Because of the advantage of being easily incorporated into existing ASV systems, spoofing detection has become an important research topic in anti-spoofing~\cite{sizov2015joint}\cite{de2012evaluation}\cite{wu2012detecting}\cite{sanchez2015}.

Many different acoustic features have been proposed to improve the performance of
Gaussian mixture model maximum-likelihood (GMM-ML) based
spoofing detection systems. In~\cite{7029029}, relative phase shift (RPS) and Mel-frequency cepstral coefficients (MFCC) were used to detect SS attacks. A fusion system combining MFCC and group delay cepstral coefficients (GDCC) was applied to resist VC spoofing in~\cite{wu2013synthetic}.
Paper~\cite{sahidullah2015} compared
the spoofing detection performance of 11 different features on the ASVspoof 2015 database~\cite{wu2015asvspoof}. Among others, dynamic linear frequency cepstral coefficients (LFCC) feature performed best on the evaluation set and the average equal error rate was lower than 1\%.

Different from the aforementioned systems, some more general systems using machine learning methods were developed to model
the difference between natural and synthetic speech more effectively.
In~\cite{xiao2015spoofing}\cite{villalba2015spoofing}\cite{chen2015robust}, spoofing detection systems based on deep neural networks (DNNs) were proposed and tested, where a DNN was used as a classifier or feature extractor. Unfortunately, experimental results showed that, compared with the acoustic feature based GMM-ML systems, these DNN systems performed slightly better on detecting the trained/known spoofing methods, but much worse on detecting unknown attacks.

In the previous studies, when a DNN was used as a feature extractor, the output of the middle hidden layer was used as DNN features to directly train some other types of models, e.g., Gaussian mixture model (GMM) or support vector machine (SVM)~\cite{villalba2015spoofing}\cite{yu2011improved}\cite{sainath2012auto}.

If we use the short-term power spectrum as the input of a DNN and set the activation function of first hidden layer as ``linear'', the learned weight matrix between the input layer and the first hidden layer can be considered as a special type of learned filter bank.  The number of this hidden layer nodes corresponds to the number of filter bank channels and each column of the weigh matrix can be treated as the frequency response of each filter. Unlike the conventional manually designed filter banks, the filters of the learned filter bank have different shapes in different channels, which can capture the discriminative characteristic between natural and synthetic speech more effectively. The DNN feature generated from the first hidden layer can be treated as a kind of filter bank feature.

Some filter bank learning methods such as LDA (Linear discriminant analysis) filter learning~\cite{burget2001data} and log Mel-scale filters learning~\cite{sainath2013learning} have been introduced in the literatures. These methods did not restrict the shapes of learned filters and the learned filter bank features were used on the speech recognition task.

%

In this paper, we introduce a new filter bank neural network (FBNN) by introducing some restriction on the training rules, the learned filters are non-negative, band-limited,
 ordered by frequencies and have restricted shapes. The DNN feature generated by the first hidden layer of FBNN has the similar physical meaning of the conventional filter bank feature and
after cepstral analysis we obtain a new type of feature, namely, deep neural network filter bank cepstral coefficients (DNN-FBCC). Experimental results show that the GMM-ML classifier based on DNN-FBCC feature outperforms the LFCC feature and DNN feature on the ASVspoof 2015 data base~\cite{wu2015asvspoof}.



\section{Filter Bank Neural Networks}
\label{sec:2}
As a hot research area, deep neural networks have been successfully used in many speech processing tasks such as speech recognition~\cite{hinton2012deep}\cite{dahl2012context}, speaker verification~\cite{variani2014deep}\cite{lee2009unsupervised} and speech enhancement~\cite{xu2014experimental}\cite{gholami2016neural}.

A trained DNN can be used for regression analysis, classification, or feature extraction. When a DNN is used as a feature extractor, due to lack of knowledge about the specific physical interpretation of the DNN feature, the learned feature can only be used to train some other models, directly. Further processing, such as cepstral analysis, can not be applied.

As one of the most classical features for speech processing, cepstral (Cep) features , e.g., MFCC and LFCC, have been widely used in most speech processing tasks.

\begin{figure}
\center
\includegraphics[width=0.45\textwidth]{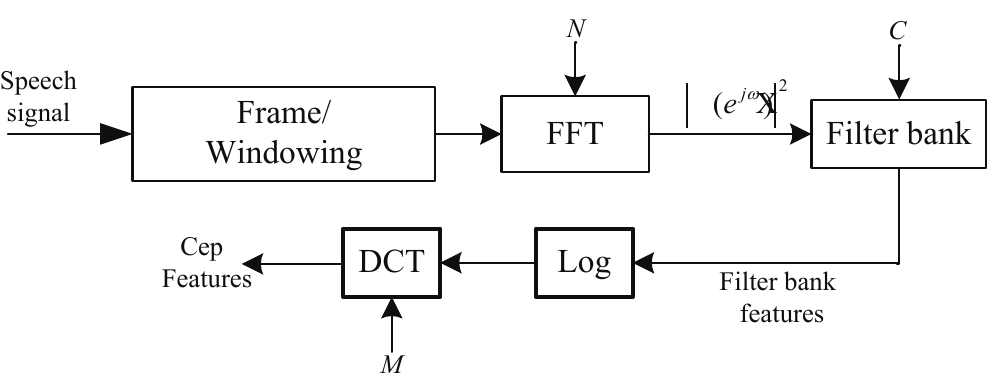}
\caption{The processing flow of computing cepstral features, where $N$, $C$, and $M$ stand for the FFT points, the number of filter bank channels, and the number of cepstral coefficients, respectively.}
\label{fig:1}
\end{figure}

Cep features can be created with the following procedure shown in Fig.\ref{fig:1}.
Firstly, the speech signal is segmented into short time frames with overlapped windows.
Secondly, the power spectrum $\left | \mathbf{X}\left( e^{jw}\right ) \right |^{2}$ are generated by frame-wise $N$ points fast Fourier transform (FFT).
Thirdly, the power spectrum is integrated using overlapping band-limited filter bank with $C$ channels, generating the filter bank features.
Finally, after logarithmic compression and discrete cosine transform (DCT) on the filter bank feature, $M$ coefficients are selected as the Cep feature.

\begin{figure}[!htp]
\centering
 \subfigure[] { \includegraphics[width=0.33\textwidth]{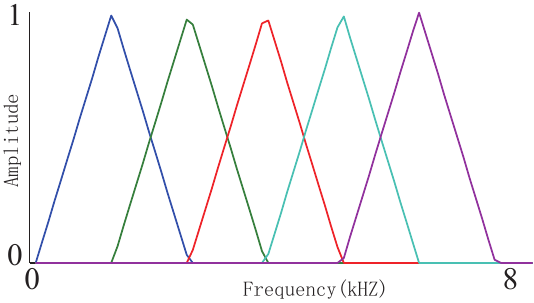} \label{fig_21} }
\subfigure[] { \includegraphics[width=0.33\textwidth]{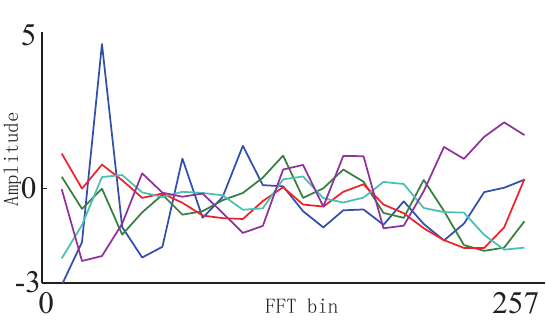} \label{fig_22} }
\subfigure[] { \includegraphics[width=0.33\textwidth]{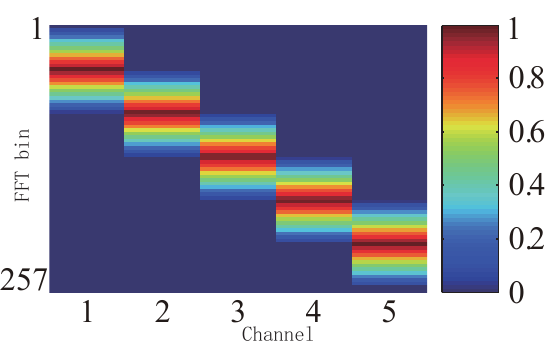} \label{fig_23} }
\subfigure[] { \includegraphics[width=0.33\textwidth]{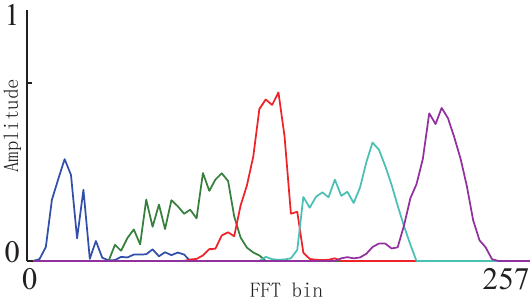} \label{fig_24} }
\caption{(a) A linear frequency triangular filter bank, (b) Learned filter bank without restriction, (c) Band-limiting mask matrix sampling from (a), (d) Learned filter bank with restriction.}
\label{fig:2}
\end{figure}

\begin{figure}[!htp]
\centering
\includegraphics[width=0.6\textwidth]{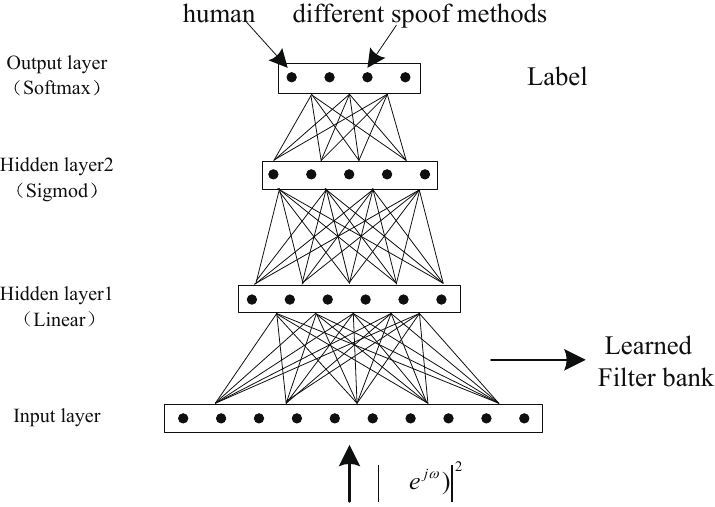}
\caption{The structure of filter bank neural networks.}
\label{fig:3}
\end{figure}

As shown in Fig.\ref{fig:2}.(a), a representative of commonly filters bank used in Cep feature extraction are non-negative, band limited, sorted by frequency and have similar shapes in different channels. The similar shapes for all the channels are not suitable for the spoofing detection task because different frequency bands may play different roles in spoofing attacks. This motivates us to use a DNN model to train a more flexible and effective filter bank.

As show in Fig.~\ref{fig:3} we build a FBNN which includes a linear hidden layer H1, a sigmoid hidden layer H2 and a softmax output layer. The number of nodes in the output layer is $N_{out}$, where the first node stands for the human voice and the other nodes represent different spoofing attack methods. The same as computing Cep features, we also use the power spectrum as the input. Because the neural activation function of H1 is a linear function, the output of the first hidden layer can be defined as:

\begin{equation}\label{eq1}
{\mathbf{H}1}=\mathbf{F}\mathbf{W_{fb}},
\end{equation}
where $\mathbf{F}$ is the input power spectrum feature with $\emph{D}$ dimension, $\emph{D}=0.5N+1$. The weight matrix between the input layer and the first hidden layer is defined as a filter bank weight matrix $\mathbf{W_{fb}}$ with dimensions $D\times C$. $C$ is the number of nodes of the first hidden layer and also means the number of channels in the learned filter bank. Each column of $\mathbf{W_{fb}}$ can be treated as a learned filter channel.

If we do not add any restrictions in the training processing, the learned filters will have the shapes as shown in Fig.~\ref{fig:2}.(b). Each channel can learn a different filter shape but the characteristics of a normal filter bank, such as non-negative, band-limit and ordered by frequency, can not be satisfied.

In order to tackle this problem, we apply some restrictive conditions on $\mathbf{W_{fb}}$ as

\begin{equation}\label{eq2}
\mathbf{W_{fb}}=\mathrm{NR}(\mathbf{W})\odot\mathbf{M_{bl}},
\end{equation}
where $\mathbf{W}\in\mathbb{R}^{D\times C}$, $\mathbf{M_{bl}}\in \mathbb{R}_{\geq 0}^{{D\times C}}$ and $\odot$ means element wise multiplication.

$\mathrm{NR}(\cdot)$ is a non-negative restriction function which can make elements of $\mathbf{W_{fb}}$ non-negative. Any monotone increasing function with non-negative output can be used. We select the sigmoid function:

\begin{equation}
\label{eq3}
\mathrm{NR}(x)=1/(1+exp(-x)).
\end{equation}
$\mathbf{M_{bl}}$ is a non-negative band-limiting shape restriction mask matrix which can restrict the filters of the learned filter bank to have limited band, regulation shape and ordered by frequency. $\mathbf{M_{bl}}$ can be generated from any band-limited filter bank by frequency-domain sampling. Fig.~\ref{fig:2}.(c) shows a $\mathbf{M_{bl}}$ sampling from a linear frequency triangular filter bank with five channels (Fig.~\ref{fig:2}.(a)).

$W_{dc}$, elements of $\mathbf{W}$, can be learned through stochastic gradient descent using equations~(\ref{eq4}) -~(\ref{eq7}):

\begin{equation}
\label{eq4}
W_{dc}=W_{dc}-\eta g_{new},
\end{equation}

\begin{equation}
\label{eq5}
g_{new}=(1-m)\times g + m\times g_{old},
\end{equation}

\begin{equation}
\label{eq6}
g=\frac{\partial L}{\partial \mathit{H1}_{c}}\frac{\partial \mathit{H1}_{c}}{\partial\mathrm W_{dc}}=\frac{\partial L}{\partial \mathit{H1}_{c}}
\mathit{F}_{d} \mathit{M_{bl_{dc}}}\frac{\partial \mathrm{NR}(W_{dc})}{\partial W_{dc}},
\end{equation}

\begin{equation}
\label{eq7}
\frac{\partial \mathrm{NR}(W_{dc})}{\partial W_{dc}}=\mathrm{NR}(W_{dc})[1-\mathrm{NR}(W_{dc})],
\end{equation}

where $d\subseteq[1,D]$, $c\subseteq[1,C]$, $\eta$ is the learning rate, $m$ is the momentum, $g$ is the gradient computed in backward pass, $g_{old}$ is the
gradient value in the previous mini-batch, and $g_{new}$ is the new gradient for
the current min-batch. $L$ is the cost function and $\frac{\partial L}{\partial \mathit{H1}_{c}}$ can be computed by
the standard back propagation equations for neural networks~\cite{rumelhart1988learning}.
The learned filters with restrictions are illustrated in Fig.~\ref{fig:2}.(d), which are band limited, ordered by frequency and have different filter shapes in different channels.

Following the cepstral analysis steps we can generate a new kind of Cep features using the filter bank generated from FBNN, which is defined as deep neural networks filter bank cepstral coefficients (DNN-FBCC). The new feature can integrate the advantages of Cep feature and the discrimination ability of DNN model, which are specially suitable for the task of spoofing detection.

\section{Experimental Results and Discussions}
\subsection{Database and Data Preparation}
The performance of spoofing detection using the DNN-FBCC feature is evaluated on the ASVspoof 2015 database~\cite{wu2015asvspoof}. As shown in TABLE~\ref{tab:AVS2015}, the database includes three sub datasets without target speaker overlap: the training set, the development set and the evaluation set. We used the training set for FBNN and human/spoof classifier training. The development set and evaluation set were used for testing.
\begin{table}[!htb]
\renewcommand{\arraystretch}{1}
\caption{{Description of ASVspoof 2015 database.}}
\label{tab:AVS2015}
\centerline{
\begin{tabular}{| c | c | c |c|c|}
\hline
& \multicolumn{2}{|c|}{Speaker} & \multicolumn{2}{|c|}{Utterances}\\
\cline{2-5}
subset &Male & Female &Genuine & Spoofed\\
\hline
Training &10 &15 & 3750 &12625\\
Development &15 &20 & 3497 &49875\\
Evaluation &20 &26 & 9404 &184000\\
\hline
\end{tabular}
}
\end{table}

Training set and development set are attacked by the same five spoofing methods, where $S1$, $S2$ and $S5$ belong to VC method and $S3$, $S4$ belong to SS method. Regarding the evaluation set, besides the five known spoofing methods, there are another five unknown methods, where $S6$-$S9$ are VC methods and $S10$ is an SS method.

The speech signals were segmented into frames with $20ms$ length and $10ms$ step size. Pre-emphasis and a hamming window were applied on the frames before the spectrum computation. Paper~\cite{sahidullah2015} showed that all the frames of speech are useful for spoofing detection, so we did not apply any voice activity detection method.
\vspace{-4mm}
\subsection{FBNN Training}
\label{subset_FBNN}
The FBNN described in Section~\ref{sec:2} was built and trained with computational network toolkit (CNTK)~\cite{yu2014introduction}.

The output layer has five nodes, the first one is for human speech and the other four are for five known spoofing methods (S3 and S4 use the same label). The number of nodes in hidden layer H2 is set as 100, the cross entropy function was selected as the cost function $L$ and the training epoch was chosen as 30. The mini-batch size was set as 128. $\mathbf{W}$ was initialized
with uniform random numbers. $\eta$ and $m$ are set as 0.1 and 0 in the first epoch, 1 and 0.9 in the other epochs.

Some experimental results published in paper~\cite{yu2016effect} and~\cite{sahidullah2015}, show that the high frequency spectrum of speech is more effective for synthetic detection.
In order to investigate the affect of different band-limiting and shape restrictions to the learned filter banks, we use four different manually designed filter banks to generate $\mathbf{M_{bl}}$: the linear frequency  triangular filter bank (TFB) with 20 channels, the linear frequency  rectangular filter bank (RFB) with 20 channels, the equivalent rectangular bandwidth (ERB) space Gammatone filter bank (GFB) with 128 channels, and the inverted ERB space Gammatone filter bank (IGFB) with 128 channels, according to the recommended in paper~\cite{adiga2013gammatone}~\cite{sahidullah2015}.

Correspondingly, the number of nodes in the first
hidden layer were set as 20, 20, 128, 128 for TFB, RFB, GFB and IGFB, respectively.
When using TFB and RFB, the dimension of the input
power spectrum is 257. The feature dimension is 513 when using GFB and IGFB.

\begin{figure}
\centering
\subfigure[] { \includegraphics[width=0.22\textwidth]{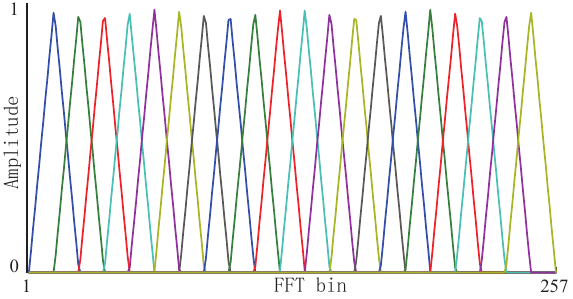} \label{fig_411} }
\subfigure[] { \includegraphics[width=0.22\textwidth]{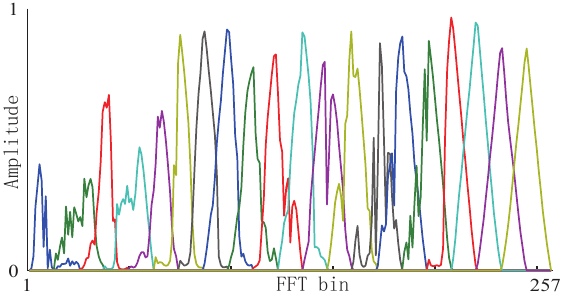} \label{fig_412} }
\subfigure[] { \includegraphics[width=0.22\textwidth]{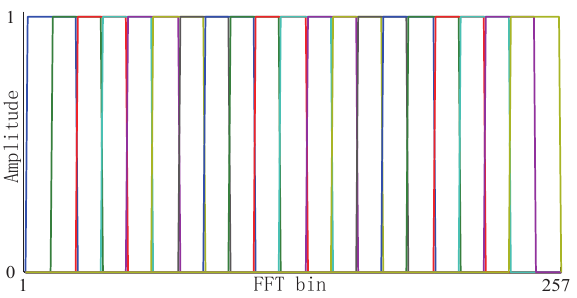} \label{fig_421} }
\subfigure[] { \includegraphics[width=0.22\textwidth]{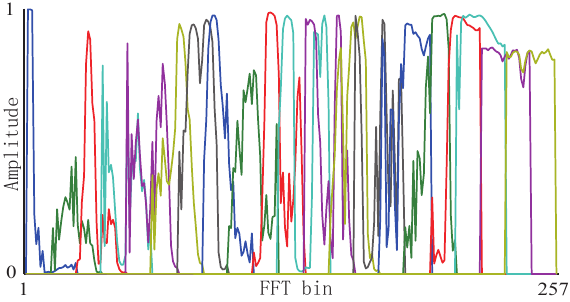} \label{fig_422} }
\subfigure[] { \includegraphics[width=0.44\textwidth]{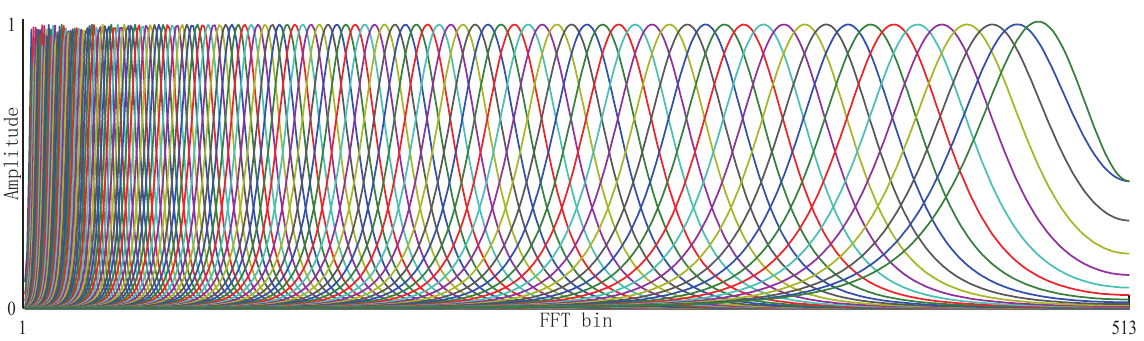} \label{fig_431} }
\subfigure[] { \includegraphics[width=0.44\textwidth]{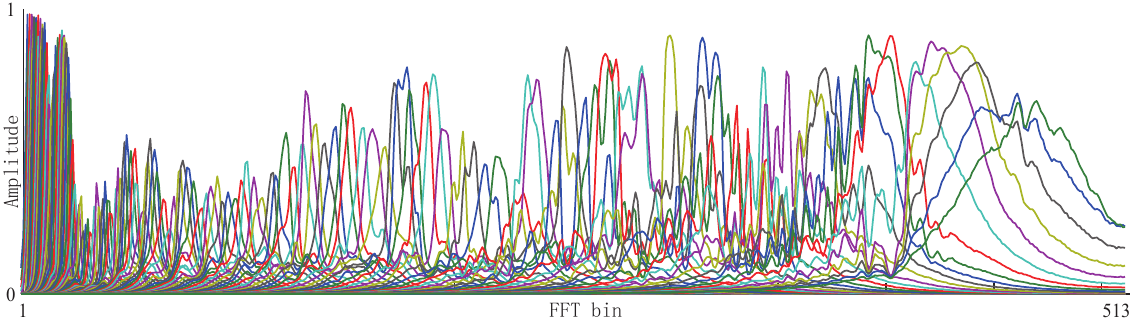} \label{fig_432} }
\subfigure[] { \includegraphics[width=0.44\textwidth]{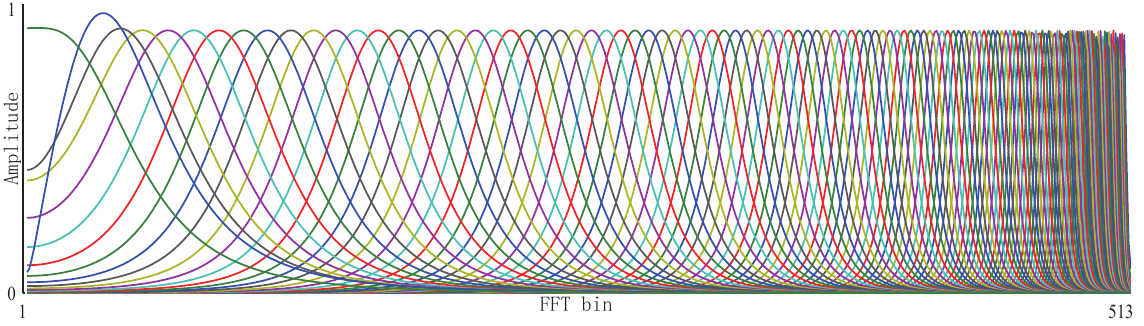} \label{fig_441} }
\subfigure[] { \includegraphics[width=0.44\textwidth]{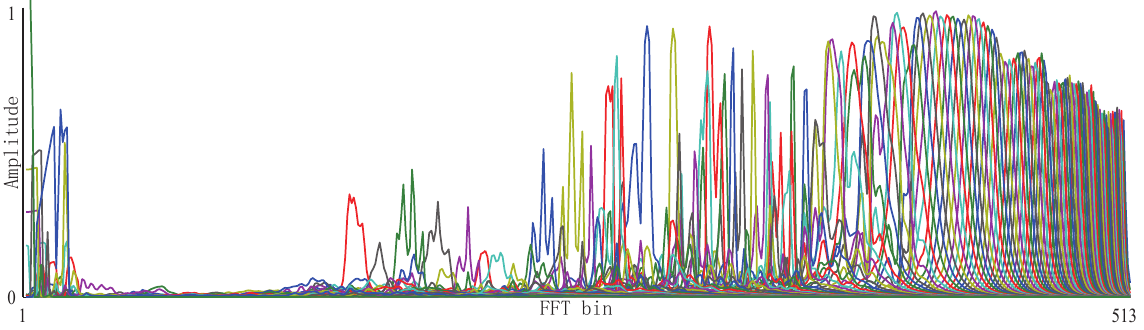} \label{fig_442} }
\caption{Filter banks used for generated $\mathbf{M_{bl}}$ and corresponding learned filter banks, (a) TFB, (b) DNN-TFB, (c) RFB, (d) DNN-RFB, (e) GFB,  (f) DNN-GFB, (g) IGFB and (h) DNN-IGFB.}
\label{fig:4}
\end{figure}

TFB and RFB equally distribute on the whole frequency region (Fig.\ref{fig_412} and Fig.\ref{fig_422}).
GFB which has been successfully used in audio recognition~\cite{adiga2013gammatone}\cite{valero2012gammatone}, has denser spacing in the low-frequency region (Fig.\ref{fig_431}) and IGFB gives higher emphasis to the higher frequency region(Fig.\ref{fig_432}).

As shown in Fig.\ref{fig:4}, after training we can get the DNN-triangle filter bank
(DNN-TFB), the DNN-rectangle filter bank (DNN-RFB),
the DNN-Gammatone filter bank (DNN-GFB) and the DNN-inverted Gammatone filter bank(DNN-IGFB). The learned filters
have flexible shapes in different frequency bands which can
capture the difference between human and spoofed speech
more effectively.

\subsection{Classifier}
In designing the classifier, we train two separated GMMs with 512 mixtures to model natural and spoofed speech, respectively. Log likelihood ratio is used as criterion of assessment, which is defined as:

\begin{equation}\label{eq6}
\mathbf{ML}\left ( \textbf{X} \right )=\frac{1}{T}\sum_{i=1}^{T}\left\{\mathrm{log}\mathrm{P}(\mathbf{X}_{i}|\lambda _{\mathrm{human}} )-\mathrm{log}\mathrm{P}(\mathbf{X}_{i}|\lambda _{\mathrm{spoof}})\right\},
\end{equation}
where $\textbf{X}$ denotes feature vectors with $T$ frames, $\lambda _{\mathrm{human}}$ and $\lambda _{\mathrm{spoof}}$ are the GMM parameters of human and spoof model, respectively.

\subsection{Results and Discussions}
We compare the spoofing detection performance between four manually designed Cep features and four DNN-FBCC features.

\begin{table}[!htb]
\renewcommand{\arraystretch}{1}
\caption{{ Description of manually designed Cep features and DNN-FBCC features used in the experiments.}}
\label{tab:featuer}
\centerline{
\begin{tabular}{|c| c | c | c |c|c|}
\hline
  & Feature&FFT &Channel &Coef. & Filter \\
  &  Name  &($N$) & ($C$) &($M$) & bank\\
 \hline
                 & LFCC &512 &20 & 20 &TFB\\
 Manually  & RFCC &512 &20 & 20 &RFB\\
 designed      & GFCC &1024 &128 & 20 &GFB\\
 Cep fearure        & iGFCC &1024 &128 & 20 &IGFB\\
\hline
          & DNN-LFCC &512 &20 & 20 &DNN-TFB\\
          & DNN-RFCC &512 &20 & 20 &DNN-RFB\\
DNN-FBCC  & DNN-GFCC &1024&128& 20 &DNN-GFB\\
          & DNN-IGFCC &1024&128& 20 &DNN-IGFB\\
\hline
\end{tabular}
}
\end{table}

As shown in Table~\ref{tab:featuer}, manually designed Cep features: LFCC, RFCC (linear frequency rectangle filter bank cepstral coefficients), GFCC (ERB space Gammatone filter bank cepstral coefficients) and IGFCC (inverted ERB space Gammatone filter bank cepstral coefficients) are generated by manually designed filter bank TFB, RFB, GFB and IGFB described in Section~\ref{subset_FBNN}.
Four DNN-FBCC features, DNN-LFCC, DNN-RFCC, DNN-GFCC and DNN-IGFCC are generated by learned filter banks DNN-TFB, DNN-RFB, DNN-GFB and DNN-IGFB, respectively.
The number of  coefficients $M$ of all the eight features are set as 20 (including the 0'th coefficient).

Inspired by the work in~\cite{sahidullah2015}, we use $\Delta$ and $\Delta^{2}$ (first- and second-order frame-to-frame difference) coefficients to train the GMM-ML classifier.
Equal error rate (EER) is used for measuring spoofing detection performance. The average EERs of different spoofing methods on development and evaluation set are shown in TABLE~\ref{tab:EERResult}.

We first conduct experiments on four manually designed Cep features, among which, IGFCC($\Delta\Delta^{2}$) performs best on detecting both known and unknown attacks and GFCC($\Delta\Delta^{2}$) works worst.
It can be inferred that the filter banks, which give higher emphasis to the higher frequency region, are more suitable for the spoofing detection task. This is inline with the finding in paper~\cite{yu2016effect}.

Then we investigate the performance of four DNN-FBCC features. DNN-RFCC($\Delta\Delta^{2}$) performs best on detecting known attacks, but works worse on unknown spoofing attacks.
This phenomena shows that the shape restrictions applied on FBNN affect the performance of spoofing detection.
When a rectangle filter is selected (RFB, Fig.\ref{fig_422}), there are no special shape restrictions
on the learned filters, and this make the learned DNN-RFCC($\Delta\Delta^{2}$) over-fits the trained/known attacks.
When a Gammatone filter is chosen (IGFB, Fig.\ref{fig_432}), the shape restriction can make the performance of DNN-IGFCC($\Delta\Delta^{2}$) better than the corresponding IGFCC($\Delta\Delta^{2}$) on both known and unknown attacks.

In general, among the eight investigated Cep features, DNN-IGFCC($\Delta\Delta^{2}$), generated by the learned filter bank which has denser spacing in the high frequency region and has the Gammatone shape restriction, performs best on ASVspoof 2015 data base and gets the best average accuracy, overall.

\begin{table}[t]
\renewcommand{\arraystretch}{1}
\caption{{ Accuracies (Avg.EER in \%) of different features on the development and evaluation set.}}
\label{tab:EERResult}
\centerline{
\begin{tabular}{|c|c | c |c|c|}
\hline
              &Dev. &\multicolumn{3}{|c|}{Eva.} \\
\cline{2-5}
Feature(dim)  & Known&  Known& Unknown & All \\
\hline
LFCC($\Delta\Delta^{2}$)(40)&0.11&0.10& 1.73 &0.92\\
RFCC($\Delta\Delta^{2}$)(40) &0.20&0.13& 1.98 &1.06\\
GFCC($\Delta\Delta^{2}$)(40) &0.74&0.48& 5.22 &2.85\\
IGFCC($\Delta\Delta^{2}$)(40) &0.13&0.07& 1.49 &0.78\\
\hline
DNN-LFCC($\Delta\Delta^{2}$)(40) &0.16&0.14&1.53&0.84\\
DNN-RFCC($\Delta\Delta^{2}$)(40) &\textbf{0.09}&\textbf{0.04}&3.01&1.52\\
DNN-GFCC($\Delta\Delta^{2}$)(40)  &0.74&0.38&4.98 &2.68\\
DNN-IGFCC($\Delta\Delta^{2}$)(40) &0.12&0.06&\textbf{1.05} &\textbf{0.56}\\
\hline
LDA-FB(20)&24.11 &23.02 &40.71&31.87 \\
DNN-BN(60)&0.22 &0.18 & 6.37 &3.28\\
l-LMFB(20)&1.79 &1.49 & 6.44 &3.96\\
DNN-BN($\Delta\Delta^{2}$)(120)&1.97 &1.46 & 4.67 &3.07\\
l-LMFB($\Delta\Delta^{2}$)(40)&0.29 & 0.18& 3.2 &1.69\\
\hline
\end{tabular}
}
\end{table}

We also compare the DNN-FBCC feature with other three data driven features which have been successfully used in speaker verification and speech recognition task: LDA filter bank feature (LDA-FB)~\cite{burget2001data}, log-normalized learned Mel-scale filter bank feature (l-LMFB)~\cite{sainath2013learning} and DNN bottle neck feature (DNN-BN)~\cite{yu2011improved}.

LDA-FB is generated by a 20 channels LDA filter bank which is learned by power spectrum feature with 257 dimension.

DNN-BN is produced by the middle hidden layer of a five hidden layers DNN, and the nodes number of hidden layers are set as 2048, 2048, 60, 2048 and 2048, respectively. The DNN is trained by a block of 11 frames of 60 MFCC(static+$\Delta\Delta^{2}$) features.

l-LMFB is generated by a neural network introduced by~\cite{sainath2013learning} which uses a 20 channel mel-scale rectangle filter bank to generate $\mathbf{M_{bl}}$ and chooses exponential function $e^x$ as a non-negative restriction function.

From the results shown in TABLE~\ref{tab:EERResult} we observe that the simple data driven filter bank feature LDA-FB is not suitable for the spoofing detection task.
Static DNN-BN, DNN-BN($\Delta\Delta^{2}$), static l-LMFB and l-LMFB($\Delta\Delta^{2}$) are all perform worse than the DNN-IGFCC($\Delta\Delta^{2}$) feature.

To sum up the learned filter banks produced by FBNN using suitable band limiting and shape restrictions can improve the spoofing detection accuracy over the existing manually designed
filter banks by learning flexible and effective filters.
DNN-FBCC, especially DNN-IGFCC($\Delta\Delta^{2}$), can largely increase the detection accuracy on unknown spoofing attacks.

\section{Conclusions}
In this paper, we introduced a filter bank neural network with two hidden layers for spoofing detection. During training, a non-negative restriction function and a band-limiting mask matrix were applied on the weight matrix between the input layer and the first hidden layer. These restrictions made the learned weight matrix non-negative, band-limited, shape restriction and ordered by frequency. The weight matrix can be used as a filter bank for cepstral analysis. Experimental results show that cepstral coefficients (Cep) features produced by the learned filter banks were able to distinguish the natural and synthetic speech more precisely and robustly than the manually designed Cep features and general DNN features.

\bibliographystyle{ieeetr}

\end{document}